\documentclass[final]{aa}  

\usepackage{graphicx}
\usepackage{txfonts}
\usepackage{lipsum}
\usepackage{subcaption}         
\usepackage{float}
\usepackage{lscape}             
\usepackage{placeins}           
\newcommand{\sevn}{{\sc Sevn}}
\newcommand{\parsec}{{\sc Parsec}}                                
\begin{document}

   \title{\bf {Reconstructing the origin of black hole mergers using sparse astrophysical models}}
\author{
V. Gayathri\inst{1}\thanks{E-mail: gayathri.v@ligo.org} \and
Giuliano Iorio\inst{2} \and
Hiromichi Tagawa\inst{3} \and
Daniel Wysocki\inst{1} \and
Jeremiah Anglin\inst{4} \and
Imre Bartos\inst{4} \and
Shubhagata Bhaumik\inst{4} \and
Zoltán Haiman\inst{5,6,7} \and
Michela Mapelli\inst{8,9,10} \and
R. O'Shaughnessy\inst{11} \and
LingQin Xue\inst{4}
}

\institute{
Leonard E. Parker Center for Gravitation, Cosmology, and Astrophysics, University of Wisconsin–Milwaukee, Milwaukee, WI 53201, USA \and
Institut de Ciències del Cosmos (ICCUB), Universitat de Barcelona (UB), c. Martí i Franquès 1, 08028 Barcelona, Spain \and
Shanghai Astronomical Observatory, Shanghai, 200030, People’s Republic of China \and
Department of Physics, University of Florida, PO Box 118440, Gainesville, FL 32611-8440, USA \and
Institute of Science and Technology Austria, Am Campus 1, Klosterneuburg 3400, Austria \and
Department of Astronomy, Columbia University, MC 5246, 538 West 120th Street, New York, NY 10027, USA \and
Department of Physics, Columbia University, MC 5255, 538 West 120th Street, New York, NY 10027, USA \and
Dipartimento di Fisica e Astronomia Galileo Galilei, Università di Padova, Vicolo dell’Osservatorio 3, I–35122 Padova, Italy \and
INFN-Padova, Via Marzolo 8, I–35131 Padova, Italy \and
INAF-Padova, Vicolo dell’Osservatorio 5, I–35122 Padova, Italy \and
Center for Computational Relativity and Gravitation, Rochester Institute of Technology, Rochester, NY 14623, USA
}




 
  \abstract
   {The astrophysical origin of binary black hole mergers discovered by LIGO and Virgo remains uncertain. 
Efforts to reconstruct the processes that lead to mergers typically rely on either astrophysical models with fixed parameters, or continuous analytical models that can be fit to observations. Given the complexity of astrophysical formation mechanisms, these methods typically cannot fully take into account model uncertainties, nor can they fully capture the underlying processes. Here, we present a merger population analysis that can take a discrete set of simulated model distributions as its input to interpret observations. The analysis can take into account multiple formation scenarios as fractional contributors to the total set of observations, and can naturally account for model uncertainties. We apply this technique to investigate the origin of black hole mergers observed by LIGO–Virgo. Specifically, we consider a model of AGN-assisted black hole merger distributions, exploring a range of AGN parameters along with several {\sc{SEVN}} population synthesis models that vary in common envelope efficiency parameter ($\alpha$) and metallicity ($Z$). We estimate the posterior distributions for AGN+SEVN models using $87$ BBH detections from the $O1--O3$ observation runs. The inferred total merger rate is $46.2\ \mathrm{Gpc}^{-3}\ \mathrm{yr}^{-1}$, with the AGN sub-population contributing $21.2\ \mathrm{Gpc}^{-3}\ \mathrm{yr}^{-1}$ and the SEVN sub-population contributing $25.0\ \mathrm{Gpc}^{-3}\ \mathrm{yr}^{-1}$.
}


   \maketitle

\section{Introduction}
By the end of the third observation run O3, the LIGO-Virgo-KAGRA \citep{TheLIGOScientific:2014jea,TheVirgo:2014hva,KAGRA:2018plz} Collaboration detected more than 90 compact binary coalescence  
\citep{LIGOScientific:2021usb,KAGRA:2021vkt}. Around $70$ of these were binary black hole (BBH) mergers with a false alarm rate (FAR) of less than one per year \citep{KAGRA:2021vkt}. Recently, the LIGO-Virgo-KAGRA collaboration released an updated detection catalog including 128 new binary events \citep{LIGOScientific:2025slb}; however, these binaries are not included in this study.  Despite this growing number of detections, the astrophysical origin of these events is uncertain. Nonetheless, these detections have led to several investigations, which includes exploring stellar evolution and binary formation mechanisms \citep{KAGRA:2021duu}, measuring cosmological parameters \citep{LIGOScientific:2021aug}, and testing General Relativity in the strong-gravity regime \citep{Ghosh:2022xhn}.

Astrophysical channels proposed for the formation of BBHs include isolated binaries, where the stars interact through processes such as mass transfer and supernova explosions, eventually forming compact objects that merge due to GW radiation \citep{LIGOScientific:2017vwq}. Another possibility is dynamical encounters, which occur in dense stellar environments like galactic nuclei and stellar clusters, where close encounters can lead to the formation of compact binaries \citep{Rodriguez:2015oxa}. 
Active Galactic Nuclei (AGN) also provide a promising site for BBH formation 
where the AGN disk can align the orbits of black holes with the disk plane, and help the merger of black holes in the disk through migration, gas capture and dynamical friction (e.g., \citealt{Bartos:2016dgn}). These channels collectively offer a comprehensive framework for understanding GW sources' observed rates and BH distributions. There are many studies to understand these distributions, few are \cite{Karathanasis:2022rtr,Colloms:2025hib,Bouffanais:2020qds,Wong:2020ise,Zevin:2020gbd}

Multiple population inference methods exist that can be used to assess the astrophysical origin(s) of the observed BBHs. These methods can be classified either as parametric or as non-parametric models. Parametric model inference assumes a typically continuous distribution for the BBH population that can be described with a small number of parameters (e.g., index of power law mass distribution), and then infers the values of these parameters \citep{2019PASA...36...10T,2019PhRvD.100d3012W,KAGRA:2021duu,LIGOScientific:2020kqk,2021ApJ...913L..23E,2022ApJ...931..108F}. 
On the other hand, non-parameter inference is typically used when the underlying astrophysical model is not a continuous function of parameters but instead is constructed, e.g., by numerical simulations. 
This type of inference is often used when a complex astrophysical merger model cannot easily be described by an analytical function and was instead obtained by simulating the system.  

\cite{Gayathri:2023bha} presented a method to extract source population information that could adopt multiple astrophysical models and carry out population inference over the ensemble of models. They introduced a mixture model method, which combined several astrophysical models with fixed parameters and one model (of AGN-assisted mergers) with a single variable parameter.  The analysis was optimized over a mixture model with different components corresponding to different formation channels. 

This work generalizes the method of \cite{Gayathri:2023bha} to incorporate a multi-dimensional parameter space in addition to a mixture model. In particular, we utilize 20 numerical simulations of AGN-assisted mergers with different environmental parameters, and the isolated binary evolution model SEVN with 19 different values of metallicity and common envelope efficiency parameter $\alpha$, and combine them to provide a sparse sampling of the highly multi-dimensional parameter space.

The paper is organised as follows. Section 2 introduces models of AGN-assisted black hole mergers and the isolated binary model for non-AGN. In Section 3, we discuss multi-model population analysis for the mixture model and the analysis setup for this study. In Section 4, we discuss the key results for different cases. In section 5, we summarise our findings and comment on future directions. 
\section{Astrophysical models}

\subsection{AGN-assisted black hole mergers} \label{sec:agn}

AGNs present a distinctive setting capable of aiding and modifying the progression of BH mergers. It is anticipated that the cores of active galaxies contain a substantial number of stellar mass BHs, potentially reaching tens of thousands, that have gravitated towards the innermost parsec via mass segregation \citep{1993ApJ...408..496M,2000ApJ...545..847M,2014ApJ...794..106A,2018Natur.556...70H,2018MNRAS.478.4030G}. The interaction between the AGN disk and these BHs can align their orbits with the disk's plane \citep{1991MNRAS.250..505S,1993ApJ...409..592A,2004ApJ...608..108G,2012MNRAS.425..460M,2014MNRAS.441..900M,2017ApJ...835..165B}. Alternatively, certain BHs may originate within the AGN disk itself. Once situated in the disk, the interplay between the swirling gas can prompt the BHs to migrate towards specific locations known as migration traps, typically around 300 Schwarzschild radii from the central supermassive black hole (SMBH). When multiple BHs reach the migration trap, they can undergo a rapid merger facilitated by dynamical friction within the disk. Conversely, BBH systems can also align their orbits with the disk, enabling them to merge swiftly within the disk before reaching the migration trap.

BH mergers facilitated by AGNs exhibit special characteristics that could set them apart from other formation pathways. These distinctions encompass their mass distribution, which is anticipated to be steeper compared to the initial black hole mass function. Additionally, their positioning within AGNs allows for differentiation from binaries formed in alternative galaxy types \citep{2017NatCo...8..831B}. 
Furthermore, BHs in AGN disks can undergo multiple consecutive mergers, giving rise to heavier black holes than what one would expect from stellar evolution only \citep{2019PhRvL.123r1101Y}.

We carried out 20 simulations of AGN-assisted black hole mergers with varying model parameters following the model presented in \cite{2021MNRAS.507.3362T}. We fixed the following parameters: supermassive black hole mass $M_{\bullet}=4 \times 10^6$\,M$_\odot$; accretion rate $0.1\dot{M}_{\rm Edd}$, where $\dot{M}_{\rm Edd}$ is the Eddington accretion rate; viscosity parameter $\alpha_{SS}=0.1$ at inner region of disc; and a radioactive efficiency $\epsilon=0.1$. We kept the initial black hole and neutron star population the same for these simulations.  We varied the following three AGN disk parameters: the ratio of the initial velocity dispersion of BHs 
to the local Keplerian velocity 
($f_{v}$), the duration of an AGN disk ($\tau \times 10^5$yr) and the inflow rate ($\lambda$) of gas from outer regions of an AGN disk 
in units of the Eddington rate.
The selected model parameters are shown in Table \ref{AGNtable}, and out of these 20 models, we discarded 6 models due to a low number of BBH samples to determine their features. Selected models are shown as $\checkmark $ in Table \ref{AGNtable}. 

\begin{table}[]
\begin{tabular}{ | p{5em} | p{1.cm}| p{1.5cm} | p{1.3cm} |p{1.cm} |} 
  \hline
Model Tag &  $f_{v}$ & Duration ($\tau$) $10^5$yr & $\lambda$ & Model included for analysis \\
  \hline \hline
 {AGN 1} & $ 0.744 $ & $ 1740.0 $& $ 0.0188 $ & \checkmark \\ \hline
 AGN 2 & $ 0.885 $ & $ 29.2 $& $ 0.0253 $  & \\ \hline
{AGN 3} & $ 0.647 $ & $ 42.1 $& $ 0.573 $ & \checkmark \\ \hline
AGN 4 & $ 0.68 $ & $ 12.4 $& $ 0.00478 $  & \\ \hline
{AGN 5} & $ 0.142 $ & $ 1000.0 $& $ 0.191 $ & \checkmark \\ \hline
AGN 6 & $ 0.936 $ & $ 13.3 $& $ 0.00102 $  & \\ \hline
{AGN 7} & $ 0.908 $ & $ 6260.0 $& $ 0.0144 $ & \checkmark \\ \hline
AGN 8 & $ 0.768 $ & $ 29.2 $& $ 0.13 $  & \\ \hline
{AGN 9} & $ 0.969 $ & $ 3020.0 $& $ 0.118 $ & \checkmark \\ \hline
AGN 10 & $ 0.376 $ & $ 49.6 $& $ 0.0455 $ & \\ \hline
{AGN 11} & $ 0.952 $ & $ 6720.0 $& $ 0.0912 $ & \checkmark \\ \hline
{AGN 12} & $ 0.866 $ & $ 1210.0 $& $ 0.398 $ & \checkmark \\ \hline
AGN 13 & $ 0.62 $ & $ 75.2 $& $ 0.0482 $  & \\ \hline
{AGN 14} & $ 0.0139 $ & $ 534.0 $& $ 0.0135 $ & \checkmark \\ \hline
{AGN 15} & $ 0.855 $ & $ 932.0 $& $ 0.012 $ & \checkmark \\ \hline
{AGN 16} & $ 0.524 $ & $ 46.6 $& $ 0.306 $ & \checkmark \\ \hline
{AGN 17} & $ 0.651 $ & $ 74.7 $& $ 0.367 $ & \checkmark \\ \hline
{AGN 18} & $ 0.694 $ & $ 813.0 $& $ 0.448 $ & \checkmark \\ \hline
{AGN 19} & $ 0.149 $ & $ 258.0 $& $ 0.0117 $ & \checkmark  \\ \hline
{AGN 20} & $ 0.743 $ & $ 4730.0 $& $ 0.00339 $ & \checkmark \\ 
\hline

\end{tabular}
\caption{AGN model parameters for our 20 simulated models.} \label{AGNtable}
\end{table}

\begin{figure*}[h]
     \centering
         \centering
\includegraphics[scale=0.37]{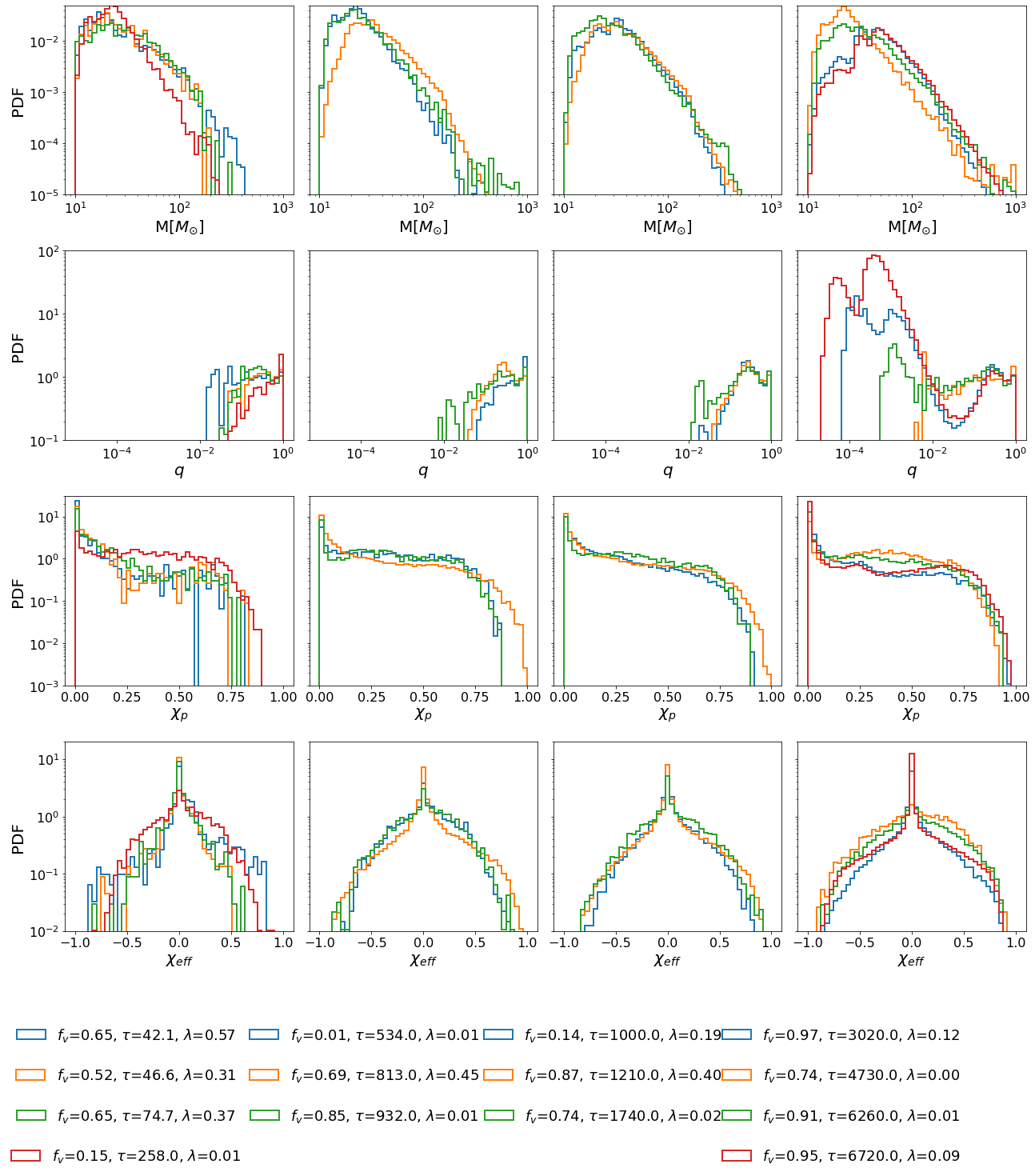}
        \caption{The probability density of AGN-assisted black hole merger properties for AGN models with different AGN parameters ($f_v$,$\tau$,$\lambda$). Panels a, b, c and d for total mass ($M(M_{\odot})$), mass ratio, $\chi_p$, and $\chi_{eff}$ parameter distributions. We have split the panels (and organized models within each panel) using the AGN duration $\tau$ (in $10^5{\rm yr}$). }
        \label{fig:agn_model_parameters}
\end{figure*}

Figure \ref{fig:agn_model_parameters} shows the BBH merger parameter distribution for our different sets of AGN model parameters. Subplots a, b, c, and d represent the total mass ($M=m_1+m_2$) in $M_{\odot}$, the mass ratio ($q=m_1/m_2$), $\chi_p$ precessing spin parameter and effective spin $\chi_{eff}$ distributions, respectively. We plot models in increasing order of $f_v$. Figure legends represent the model parameters for each column of the figure. To show the difference in binary parameters over model parameters, we have plotted limited models in each subplot. In the first and last columns, we plot 4 models; in the second and third columns, we plot 3 models.   

As the AGN disk duration ($\tau$) and velocity dispersion ($f_v$) parameters increase, we find heavier, highly asymmetric, highly-precessing spin and a 
wide 
range of effective spin binary systems formed compared to low $\tau$ AGN models. 
This is mainly 
because we have more time to form extraordinary binaries. 
For high-$\tau$ models, 
we observed a total mass distribution peak 
shifting 
towards ($50-70$ $M_{\odot}$) compared to low-$\tau$ models. 
{This could be due to a larger number of second 
and higher 
generation mergers}. The AGN disk duration parameter has the largest impact relative to the other two parameters.   
         
\begin{figure*}[hbt!]
    \includegraphics[trim={4.8cm 0.5cm 4.8cm 0cm},clip,scale=0.39]{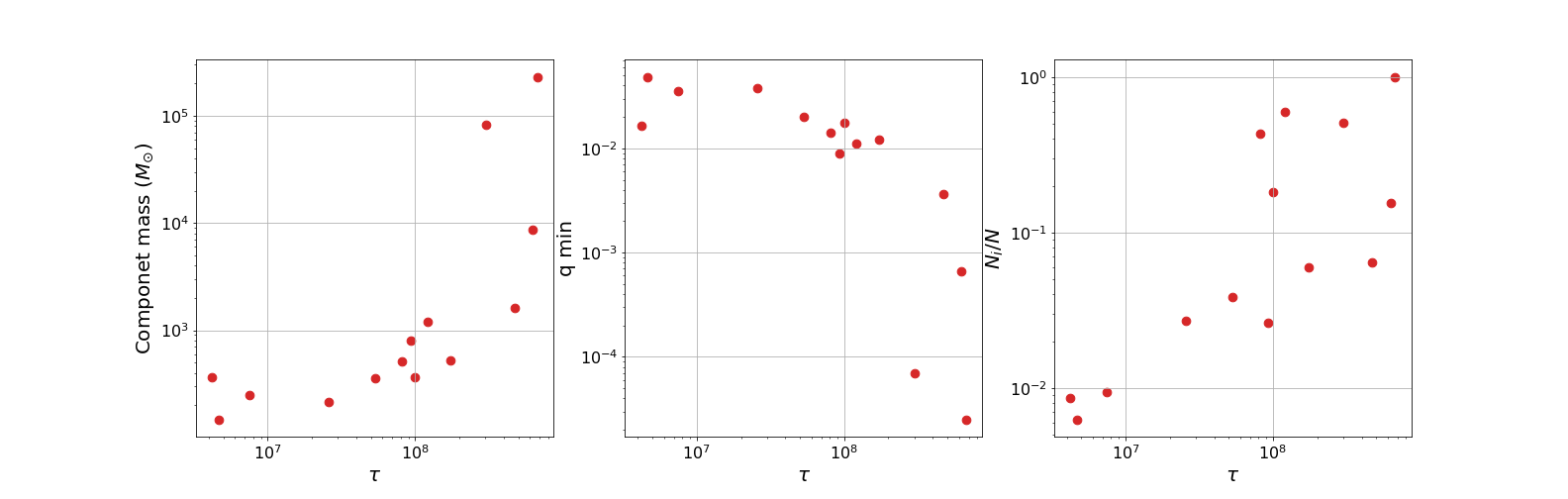}
         \caption{Extremal binaries produced in our models, versus AGN duration. Panels show the maximum mass of the component objects, the minimum mass ratio ($q_{\text{min}}$), and the fraction of binaries ($N_i/N$).}
        \label{fig:agn_duration_pro}
\end{figure*}
Figure \ref{fig:agn_duration_pro} shows the binary properties of AGNs, including the maximum mass of the component objects, the minimum mass ratio ($q_{\text{min}}$), and the fraction of binaries ($N_i/N$). Here, $N_i$ represents the number of binaries in the ${i}^{th}$ AGN model, and $N$ denotes the number of binaries in the highest $\tau$ AGN model.

\begin{figure*}[hbt!]
\includegraphics[trim={4.8cm 0cm 4.8cm 1cm},clip,scale=0.39]{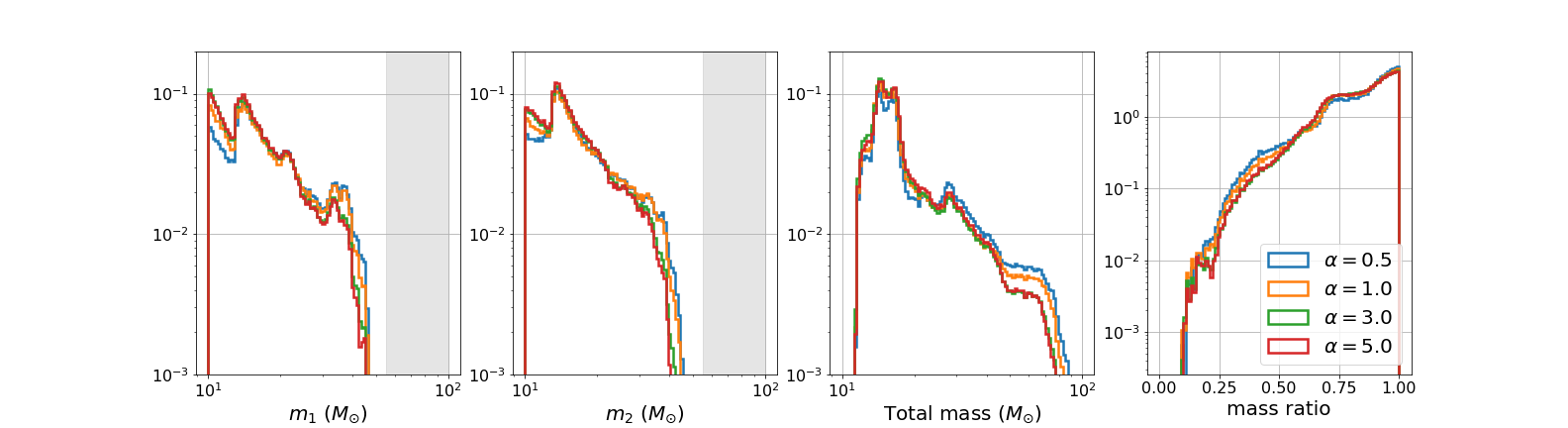}
\includegraphics[trim={4.8cm 0cm 4.8cm 1cm},clip,scale=0.39]{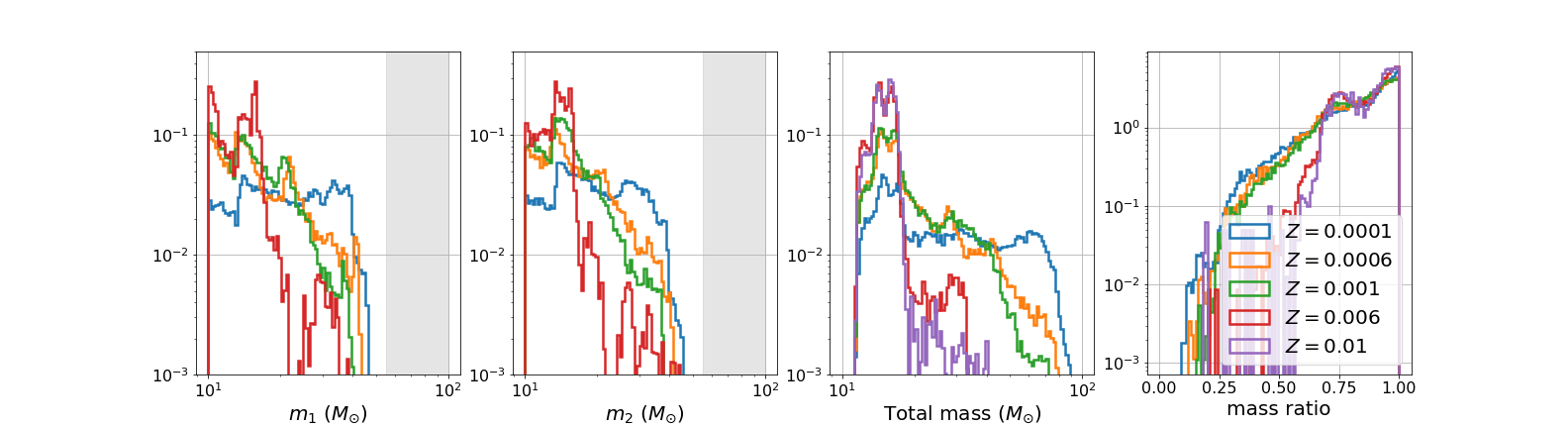}
\caption{Binary merger parameter distributions for SEVN models with different $\alpha$ and $Z$ values. The first and second columns correspond to primary and secondary masses for binaries. The third and fourth columns correspond to the total mass and mass ratio for binaries. The grey shade area represents the PISN mass gap region.}
\label{fig:SEVN-model-alpha-Z}
\end{figure*}

\subsection{Isolated binaries }\label{sec:sevn}
Stellar evolution is a complex process that involves the transformation of stars over their lifetimes, from the initial stage on the zero-age main sequence (ZAMS) to their ultimate fates, such as white dwarfs, neutron stars, or BHs. Understanding the evolution of isolated binary star systems is particularly important, as they can lead to the formation of compact objects that are in binaries from birth. 

We made use of the binary formation and evolution code SEVN (Stellar EVolution for N-body) \citep{Spera:2018wnw,2022ascl.soft06019S,Iorio:2022sgz,Mapelli:2020xeq}, which has been utilized extensively to calculate stellar evolution by interpolating pre-computed sets of stellar tracks. The advantage of SEVN lies in its flexibility and generality, allowing for easy adaptation and updates to the stellar evolution models. This adaptability is achieved by loading new sets of look-up tables. This means that the code can incorporate the latest improvements in stellar evolution research. Over the years, the SEVN code has undergone multiple revisions and improvements. This includes enhancements in time-step calculations, modularity, and the inclusion of new functionalities and options to broaden its scope and applicability. 

At the core of SEVN's calculations are the stellar-evolution tables, which provide a wealth of information about the evolution of stars based on their initial mass (MZAMS) and metallicity ($Z$). These tables track seven essential properties of stars during their evolution: time, total stellar mass, He-core mass, CO-core mass, stellar radius, bolometric luminosity, and the stellar phase. Additionally, SEVN offers the option to include other tables to explore further characteristics of stars as they progress through different stages of their evolution. We adopted the latest version of the \parsec\ stellar evolution models \citep{2012MNRAS.427..127B, 2015MNRAS.452.1068C, 2019MNRAS.485.4641C, 2022A&A...665A.126N}, as implemented in \citet{Costa:2023xsz}. A detailed description of the interpolation methods and their integration into \sevn\ is provided in \citet{Iorio:2022sgz} and \citet{Korb:2024igp}.

In the context of compact object formation, SEVN incorporates several models of core-collapse supernovae (SN) models, among which the delayed and rapid model proposed by \citet{2012ApJ...749...91F}. Core-collapse SN play a crucial role in shaping the final fate of massive stars, leading to the formation of neutron stars and BHs. By incorporating these developed models, SEVN provides a valuable view into the complex connection between binary star evolution and the formation of compact objects.

As a result, the SEVN binary population synthesis code has proven to be an important tool for studying the evolution of binary stars and the formation of compact objects. The main advantage is its ability to interpolate pre-computed stellar tracks and its flexibility in adapting to updated stellar evolution models, which results in it being a powerful choice for exploring the diverse and intriguing pathways of stellar evolution in binary systems.

With the updated version of the SEVN synthesis code, we used binary distribution from \citep{2023MNRAS.524..426I}, which used $1.2\times10^9$ binaries in the metallicity range $0.0001 \leq Z \leq 0.03$, exploring a number of models for electron-capture, core-collapse and pair-instability SN, different assumptions for the common envelope, stability of mass transfer, quasi-homogeneous evolution and stellar tides. Some peculiar LVK events even challenge current evolutionary models, indicating the existence of compact objects inside the claimed lower and upper mass gap. Assuming that there is insufficient time for the system to realign spins between the first and second SN events, and considering that SN only alter the direction of the binary system's orbital angular momentum, not the spins of the compact objects themselves, we can determine the angle between the spins of the two compact objects and the orbital angular momentum of the binary system as follows (Equation 1.16 from \citealt{Mapelli:2021taw}),
\begin{equation}
    \cos(\theta) = \cos(\nu_1) \cos(\nu_2) + \sin(\nu_1) \sin(\nu_2) \cos(\phi),
\end{equation}
where $\nu_i$ (with i = 1, 2) is the angle between the new and the old orbital
angular momentum after an SN (i = 1 corresponding to the first SN, i = 2 corresponding to the second SN), and $\phi$ is the phase of angular momentum projection into the orbital plane. For this study, we consider uniform distribution in $\phi$ and the magnitude of the spins ($a_1$ and $a_2$) generated with Maxwellian with a one-dimensional root mean square of 0.05 or 0.1.

For this study, we have considered two SEVN parameters common envelope efficiency parameter and metallicity. Figure \ref{fig:SEVN-model-alpha-Z} shows the binary merger parameter distribution from the SEVN model with different $\alpha$ and $Z$. The top panel corresponds to different $\alpha$ keeping $Z$ 
within the range 
$0.0001 \leq Z \leq 0.03$, and the bottom panel corresponds to 
different 
$Z$ keeping $ \alpha$ in the full range.  By changing the $\alpha$ value, we see the change in total mass and mass ratio, which could change how these models play a role in population analysis. In the case of changing metallicity $Z$, we have shown five cases which correspond to $Z=0.0001, 0.0006, 0.001, 0.006, 0.01$; for component masses columns, we plotted four cases (except $Z=0.01$ ) to have feature clarity. Due to our toy model, please note that we will not see much change in spin parameters $\chi_{eff}$ and $\chi_p$.

\subsection{Power-law-Peak empirical model }\label{sec:plg}
Power-Law + Peak model is a basic empirical framework used in GW population analyses to describe the distribution of 
BH 
masses in binary systems. This model has two component parts: the first one is the Power Law, and the second one is the Peak. The power-law(pl) component is characterized by a spectral index ($\alpha_{pl}$, which dictates the slope of the distribution.
This component captures the decreasing frequency of more massive black holes, consistent with a bias towards lighter black holes. The peak component is a Gaussian-like peak around a specific mass range, often referred to as the "peak" mass. The peak accounts for an excess of black holes around this mass, which may correspond to a preferred mass scale in the population. The location and width of this peak are free parameters in the model. This allows for flexibility in fitting to observational data. The model often incorporates a maximum mass cutoff, beyond which the probability of finding black holes significantly drops. This is consistent with the idea that there is an upper limit to black hole masses formed through stellar evolution. The mass ratio is drawn from another power law, and the spins are drawn from an unknown Beta distribution.

\section{Multi-model population analysis}
In this section, we describe the flexible parameter method used for the BBH population study to estimate the merger rate as a function of BBH parameters. 
\citet{LIGOScientific:2020kqk} used different sets of empirical models to describe detected BBHs. These analyses showed features in the mass spectrum around $30-40 M_{\odot}$. In our previous study \citep{Gayathri:2021xwb}, we considered mixture models of an AGN model and power-law-peak. Here, we extend our previous work to incorporate multiple AGN models, isolated binaries with different parameters, and the power-law-peak model. In this work, we have used parameter samples from O1-O3 events with approximation IMRPhenomPv2 and SEOBNRv4PHM. We use simple Gaussian fits in (${\cal{M}},\eta,\chi_{eff}$) to the events taken from this catalogue \cite{Delfavero:2021qsc}. 


\subsection{Population Inference Formalism}

We first describe the population inference method for multiple formation channel contributions, following \citet{Wysocki:2018mpo}. The binary merger detection rate is  
\begin{equation}
    {dN/dV_c dtdx={\cal{R}}p(x),}
\end{equation}
where $x$ is a set of intrinsic binary parameters (such as mass $m_i$ and spin ${\bf S}_i$ for $i=1,2$ ), $N$ is the total number of detections, $V_c$ is the comoving volume, $\cal{R}$ is the space-time-independent rate of binary coalescence per unit comoving volume and $p(x)$  is the probability density of having $x$ parameters for a detected binary. The likelihood of the  astrophysical BBH population having a
given merger rate is 
\begin{equation}
    {\cal{L}}({\cal{R}},X)  \equiv  p({\cal{D}}|{\cal{R}},X)
\end{equation}
where $X\equiv (m_1,m_2,\chi_1,\chi_2)$, $\chi_i=|{\bf S}_i|/m_i^2$, and ${\cal{D}}={d_1,d_2,...,d_n}$ is the data for $N$ detections. We can rewrite the above equation as

\begin{equation}
    {\cal{L}}({\cal{R}},X) \propto e^{-\mu({\cal{R}},X)} \prod_{n=1}^N \int dx p(d_n,x) {\cal{R}} p(x,X).
\end{equation}
Here, $\mu({\cal{R}},X)$ is the expected number of detections under a given population parametrization $X$ with overall rate $\cal{R}$. By applying Bayes' theorem, one can derive a posterior distribution for 
$R$ and $X$ after assuming a prior $p(R,X)$.

\subsection{Multi-model merger rate}
Our model is a multi-formation model, parameterised by the unknown (continuous) AGN merger rate and its (discrete) AGN parameters, parameterised by the unknown (continuous) stellar evolution merger rate and its parameters, along with an empirical mixture model.  The overall merger rate density $dN/dVdXdt$ can, therefore, be expressed as a sum

\begin{align}
\frac{dN}{dVdXdt} & \, = 
{\cal R}_{\rm agn} p_{agn}(X|\Lambda_{agn}) + {\cal R}_{\rm sevn} p_{sevn}(X|\Lambda_{sev}) +  \\
&  \text{\quad} 
\begin{aligned} 
 \text{\quad}{\cal R}_{g}p_g(X|\Lambda_g) + {\cal R}_{pl}p_{pl}(X|\Lambda_{pl})\\ 
\end{aligned}
\end{align}
where $X$ are binary parameters and where $p_q,\Lambda_q$ are the model distributions and parameters for the $q$th component (AGN, SEVN, Gaussian, and power-law, respectively). Using the above formalism, we estimate the fraction of BBHs that are produced through the AGN, isolated binary, as well as empirical models based on detected GW merger data. For this study, we enhanced existing parametric methods by incorporating a mixture model with a larger parameter space. This allows us to estimate model parameters which represent detected GW events.
\subsection{Analysis set-up}

In this analysis, we made use of BBH mergers detected by LIGO/Virgo during their first three observing runs: O1, O2, and O3 \citep{2018arXiv181112907T,2020arXiv201014527A,LIGOScientific:2021usb,KAGRA:2021vkt}. We adopted these mergers' reconstructed mass, spin and distance parameters from the GWTC open data centre. Additionally, we incorporated 8 more events which have been detected in a deep reanalysis of O3a data \cite{LIGOScientific:2021usb}.  In the case of the astrophysical source population, we considered  $14$ AGN models, $15$ SEVN models with different $Z$ and $4$ SEVN models with different $\alpha$. While considering the $Z$ parameter, we do not constrain $\alpha$ and vice versa. In this study, we also included some empirical models along with AGN models, as well as SEVN models. We used a power-law plus peak and a broken power law for the empirical model cases.   

\section{Results}
Here we discuss our findings with different population models and their merger rates. 
\subsection{Binary model: AGN models only}
We used the suite of 14 AGN models for the multi-model analysis, as discussed in Section \ref{sec:agn}, noting that the analysis can be carried out on virtually any number of simulations. For this first analysis, we assumed that all detected BBHs were from AGN origin.   

\begin{figure}[hbt!]
\includegraphics[width=0.55\textwidth]{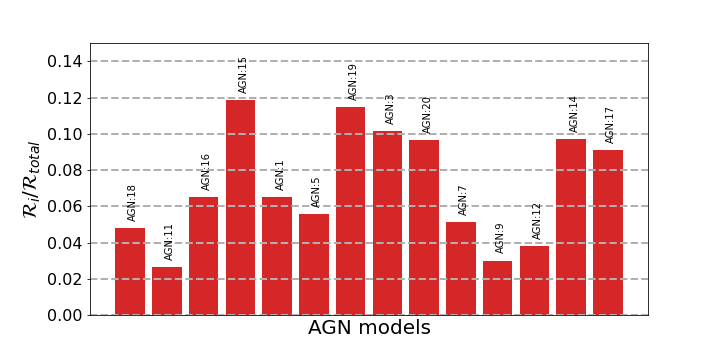}

\caption{The merger rate fraction for AGN models with 87 BBH detections from O1-O3 observation runs. Each bar corresponds to one AGN model. } 
\label{fig:merger_rate_bar_agn_only}
\end{figure}

Figure \ref{fig:merger_rate_bar_agn_only} shows the estimated merger rate for BBHs via AGN formation channels for each AGN model. When all 14 AGN models are combined into a single model, the estimated rate is $39.5^{+10.34}_{-11.06},{\rm Gpc}^{-3},{\rm yr}^{-1}$. Alternatively, if the models are treated individually, their separate contributions can be determined.  In that case, 2 models have a higher rate than others, which shows that they contribute more to the overall rate. These models are AGN:15 and AGN:19, which correspond to $(f_v,\tau, \lambda)$: $(0.855, 932.0, 0.012)$ and $(0.149, 258.0, 0.0117)$ AGN model parameters.  Out of $14$ models, six models show a good representation of detected BBHs. 

We translated the rate estimation into AGN model parameter estimation, the estimated $f_v$ is 
$0.86^{+0.05}_{-0.71}$, the AGN duration $\tau$ supports more on lower value 
$ 42.1 \times 10^{5}$ and the inflow gas rate 
$\lambda$ is to be ${0.31}^{+0.09}_{-0.29}$, these values are estimated in $65\%$ creditable interval. These values have large variability because we have estimated these parameters based on $14$ discrete model values. 

\subsection{Binary model: SEVN models only}

We use only SEVN models, divided into two subcategories based on common envelope efficiency parameter ($\alpha$) and metallicity ($Z$). The 
$\alpha$ parameter includes four values, while metallicity spans $15$ distinct values. We perform our analysis using these model subcategories, assuming all $87$ detected events originate from stellar evolution. Our goal is to identify which models best support these events and to examine how the merger rate varies accordingly.

\begin{figure}[hbt!]
\includegraphics[width=0.5\textwidth]{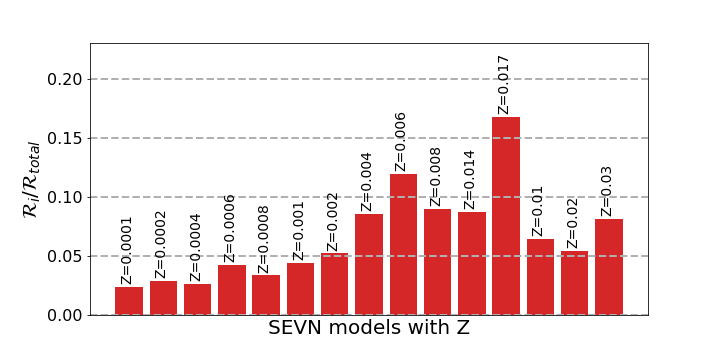}

\caption{The merger rate fraction for SEVN models with different $Z$. Each bar corresponds to one $Z$ value.} 
\label{fig:merger_rate_bar_sevn_z_only}
\end{figure}

For 
models with different $\alpha$, we have estimated the rate fraction for each model. All four models contribute similarly to the rate, ranging in  
$17 \% - 27\% $, with the highest 
contribution
for $\alpha=1$ and the lowest for $\alpha=0.5$. We 
did not observe 
any specific favor for a particular $\alpha$ value. But that is not the case for models categorized by metallicity value. It supports 
higher contributions from medium- to high-metallicity models, 
as shown in Figure~\ref{fig:merger_rate_bar_sevn_z_only}. Its fractional contribution varies from $\sim 3\%$ to $\sim 17\%$, with the lowest contribution at $Z=0.0001$ and the highest at $Z=0.017$.       
The population being studied, observed BBHs, matches best with the predictions of the stellar evolution model that assumes solar-like metallicity. 

\subsection{Binary model: SEVN models plus AGN models}

In a third analysis, we combined AGN models and SEVN models to characterize the origin of the 87 detected BBHs. Our primary objective is to incorporate the stellar evolution aspect of the population, as described by SEVN models, and complement it with various non-stellar models based on AGN models with different parameter settings.

\begin{figure*}[!tbp]
  \begin{minipage}[b]{0.49\textwidth}
    \includegraphics[width=\textwidth]{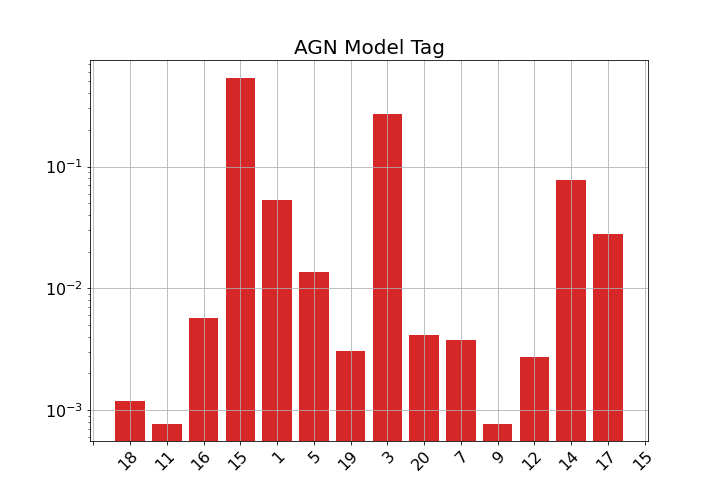}
  \end{minipage}
  \hfill
  \begin{minipage}[b]{0.49\textwidth}
    \includegraphics[width=\textwidth]{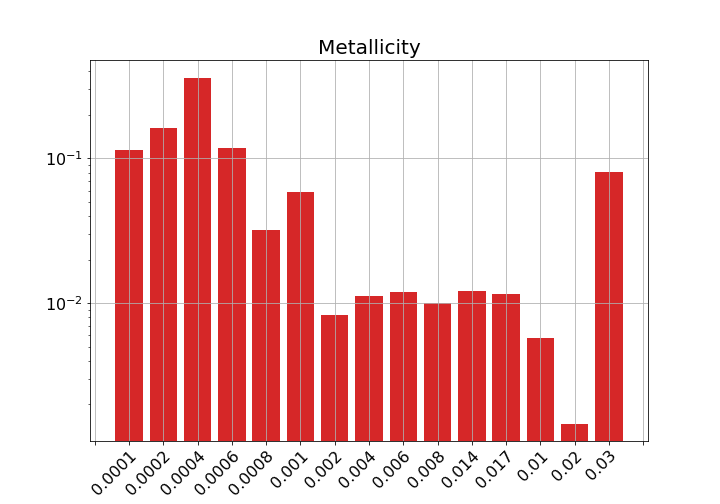}
  \end{minipage}
  \caption{The merger rate estimation for SEVN and AGN with 87 BBH detections from O1-O3 observation runs. The left figure shows different AGN models, and the x-axis represents the AGN model tag. The right figure for different SEVN models, and the x-axis represents the metallicity of SEVN models.}
\label{fig:rate_from_agn_sevn}
\end{figure*}

Figure \ref{fig:rate_from_agn_sevn} illustrates the contributions of each model to a specific sub-population. The left panel displays the AGN sub-population, while the right panel represents the SEVN sub-population. We have observed that the AGN model parameter estimation is consistent if we add the SEVN model to this study. The estimated values are  $f_v=0.09^{0.21}_{-0.08}$, the AGN duration is $932 \times 10^{5}$ years and the inflow gas rate is to be ${0.74}^{+0.21}_{-0.22}$. Along with that, we also estimate the SEVN model parameter $Z$, which peaks around $Z=6 \times 10^{-4}$.

\begin{figure*}[hbt!]
\includegraphics[width=\textwidth]{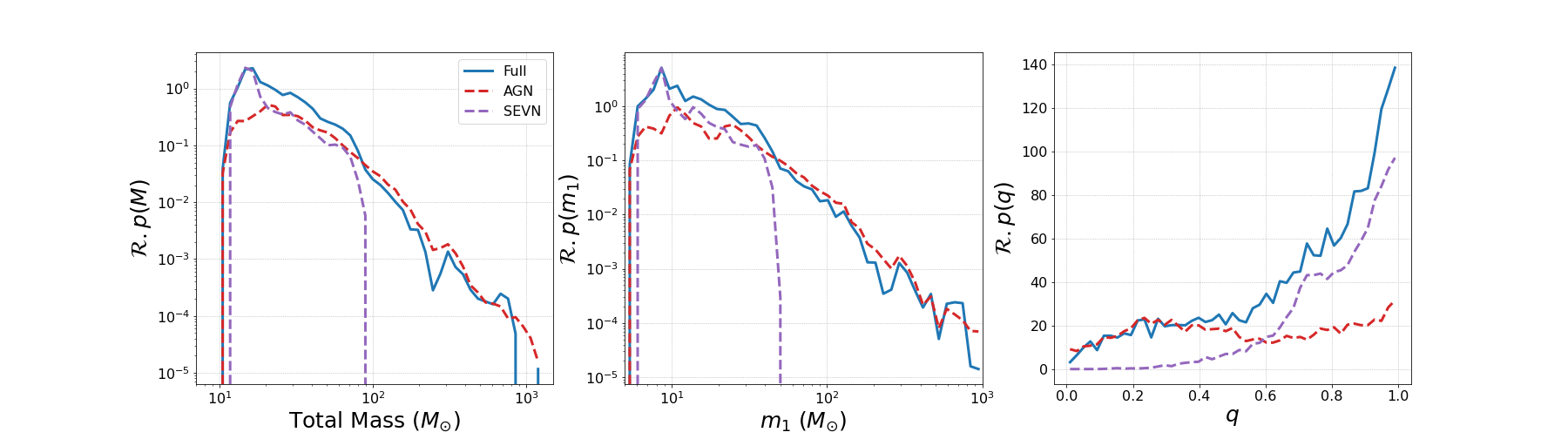}
\caption{The posterior distribution estimation for AGN+SEVN models with 87 BBH detections from O1-O3 observation runs. The total estimated rate is $46.26$ ${\rm Gpc}^{-3}\,{\rm yr}^{-1}$, the AGN sub-population rate is $21.21$ ${\rm Gpc}^{-3}\,{\rm yr}^{-1}$, and the SEVN sub-population rate is $25.06$ ${\rm Gpc}^{-3}\,{\rm yr}^{-1}$ }
\label{fig:merger_posterior_distribution_agn_sevn}
\end{figure*}

Figure \ref{fig:merger_posterior_distribution_agn_sevn} depicts the posterior distribution for each sub-population in comparison to the overall distribution. AGN models predominantly influence the low mass ratio region, while SEVN models dominate the high mass region. In the context of an asymmetric mass system, AGN models tend to favour low mass ratio systems (low $q$) instead of systems with equal masses ($q=1$). Conversely, in the case of the SEVN model, it exhibits a stronger preference for equal mass systems over low $q$ systems.  

\begin{figure*}[hbt!]
\includegraphics[width=\textwidth]{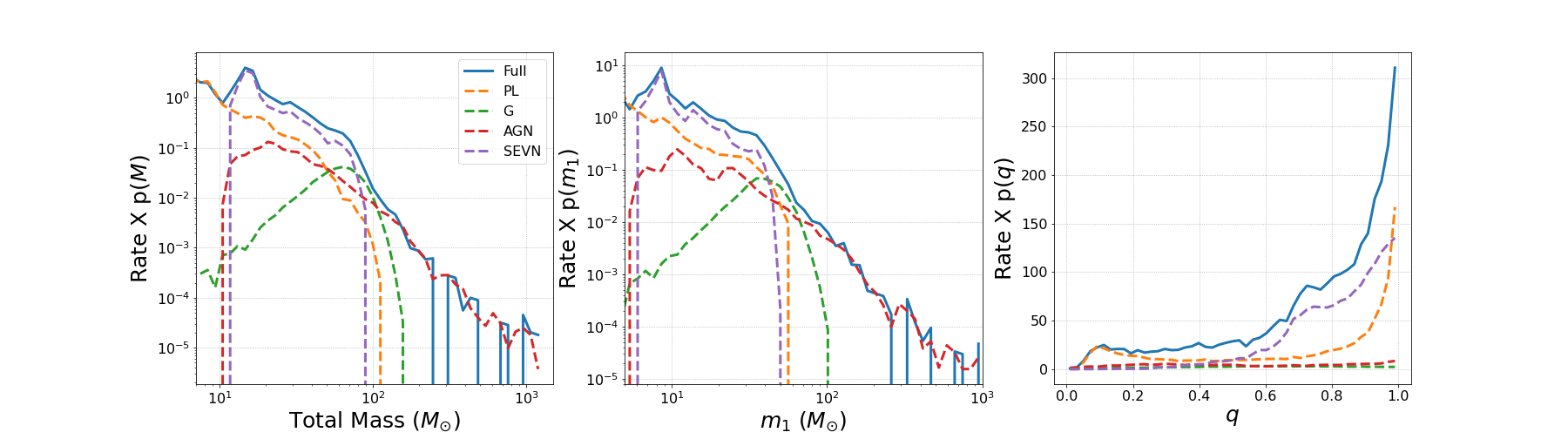}

\caption{The merger rate estimation for PL+G+AGN+SEVN models with 87 BBH detection from O1-O3 observation runs. The total estimated rate is $58.1$ ${\rm Gpc}^{-3}\,{\rm yr}^{-1}$, the PL sub-population rate is  $29.56$ ${\rm Gpc}^{-3}\,{\rm yr}^{-1}$, the 
Gaussian sub-population rate $2.66$ ${\rm Gpc}^{-3}\,{\rm yr}^{-1}$, the AGN sub-population rate is $5.1$ ${\rm Gpc}^{-3}\,{\rm yr}^{-1}$, and the SEVN sub-population rate is $31.63$ ${\rm Gpc}^{-3}\,{\rm yr}^{-1}$. 
}
\label{fig:posterior_plg_agn_s}
\end{figure*}

Figure \ref{fig:posterior_plg_agn_s} shows the posterior distributions of each sub-population compared to the overall distribution for the power-law (PL) plus Gaussian (G) model \citep{LIGOScientific:2020kqk}, along with the AGN and SEVN models. We employ the standard power-law plus peak model, previously used in our earlier studies \cite{Gayathri:2023bha,Gayathri:2021xwb}, which represents stellar-evolution binary populations. In this analysis, we include two stellar-evolution models to cover the full range of stellar-origin black holes. The results indicate that both stellar models predominantly occupy the lower-mass and near-equal mass-ratio parameter space. In contrast, the AGN model continues to dominate as a standalone model that supports higher-mass regions of the parameter space.  

\section{Conclusion}
This paper addresses the astrophysical origins of BBH mergers detected by the LIGO-Virgo collaborations. By comparing AGN and SEVN BBH formation models with the reconstructed sample of observed BBHs, we analyze each model characterized by specific parameters. For the AGN model, the parameters include the initial velocity dispersion of black holes, the duration of an AGN disk, and the inflow rate of gas from the outer regions of an AGN disk. For the SEVN model, the parameters are metallicity and the common envelope efficiency parameter, $\alpha$.  

We estimate the overall merger rate density for BBH sources from these models and assess individual formation channel contributions. Our findings indicate that the AGN model contributes across the entire mass and mass ratio spectrum, particularly dominating the high mass and low mass ratio regions. Conversely, the SEVN model shows a higher contribution in the low mass range and for systems with nearly equal masses. Additionally, we employ Bayesian analysis to estimate the limits for the model parameters $f_v$, $\tau$, $\lambda$, $Z$, and $\alpha$. 

Our study provides a comprehensive analysis of BBH formation scenarios, highlighting the significant roles of both AGN and SEVN models in explaining the observed BBH mergers, and offering insights into the parameter limits that govern these astrophysical processes.

In this study, we employed a suite of 14 AGN models to estimate the merger rate of BBHs detected during the O1-O3 observation runs. By assuming that all detected BBHs originated from AGN formation channels, we estimated the AGN merger rate to be \(39.5^{+10.34}_{-11.06}\, \text{Gpc}^{-3} \text{yr}^{-1}\), with a 65\% credible interval. Among the AGN models, four of them contributed more than 10\% to the overall rate, with AGN:15 offering the highest support. The estimated values are  $f_v=0.09^{0.21}_{-0.08}$, the AGN duration is $932 \times 10^{5}$ years and the inflow gas rate is to be ${0.74}^{+0.21}_{-0.22}$. Along with that we also estimate SEVN model parameter $Z$, which peaks around $Z=6 \times 10^{-4}$.

In a second analysis, we integrated AGN models with SEVN models to account for stellar evolution and its role in BBH formation. This combined approach provided consistent AGN parameter estimates, showing distinct contributions from AGN and SEVN models. Specifically, AGN models predominantly influenced low mass-ratio systems (low $q$), while SEVN models favored equal-mass systems ($q = 1$). This multimodel analysis improves our understanding of BBH origins, offering valuable insights into their formation channels and mass distribution.

The fourth observation run (O4) is currently ongoing, and we expect a significant increase in detections by the end of this run. In our future studies, we plan to expand the set of AGN models to cover a broader range of AGN parameter space. Incorporating O4 detections along with these extended AGN models will enable us to constrain AGN parameters and their overall contribution with greater accuracy. Notably, we have already identified a very high-mass BBH (GW231123) \cite{2025arXiv250708219T}, which will help reduce uncertainties in the high-mass region.
    
\vspace{1cm}
\noindent{\bf Acknowledgements}
We gratefully acknowledge the support of LIGO and Virgo for the provision of computational resources. G.V. and D.W. acknowledge the support of the National Science Foundation under grant PHY-2207728.  I.B. acknowledges the support of the National Science Foundation under grants \#1911796, \#2110060 and \#2207661 and of the Alfred P. Sloan Foundation. 
G.I. was supported by a fellowship grant from the la Caixa Foundation (ID 100010434). The fellowship code is LCF/BQ/PI24/12040020.
ROS acknowledges support from NSF PHY 2309172 and 2012057.
Z.H. acknowledges support from NASA grants 80NSSC22K0822 and 80NSSC24K0440.
H.T. acknowledges support from the Center for Computational Astrophysics at the National Astronomical Observatory of Japan for providing numerical resources on Cray XD2000. 
This research has made
use of data, software and/or web tools obtained from
the Gravitational Wave Open Science Center (https:
//www.gw-openscience.org), a service of LIGO Laboratory, the LIGO Scientific Collaboration and the Virgo
Collaboration.
LIGO is funded by the U.S. National Science Foundation. Virgo is funded by the French Centre National de Recherche Scientifique (CNRS), the Italian Istituto Nazionale della Fisica Nucleare (INFN) and the
Dutch Nikhef, with contributions by Polish and Hungarian institutes. This material is based upon work supported by NSF's LIGO Laboratory, which is a major facility fully funded by the National Science Foundation.

\bibliographystyle{aa}   
\bibliography{paper}

\end{document}